\documentclass{article}
\usepackage{spconf,amsmath,graphicx,hyperref}
\usepackage{cite}
\usepackage{amsmath,amssymb,amsfonts}
\usepackage{algorithmic}
\usepackage{graphicx}
\usepackage{textcomp}
\usepackage{xcolor}

\usepackage[utf8]{inputenc} % allow utf-8 input
\usepackage[T1]{fontenc}    % use 8-bit T1 fonts
\usepackage{hyperref}       % hyperlinks
\usepackage{url}            % simple URL typesetting
\usepackage{booktabs}       % professional-quality tables
\usepackage{amsfonts}       % blackboard math symbols
\usepackage{nicefrac}       % compact symbols for 1/2, etc.
\usepackage{microtype}      % microtypography
\usepackage[dvipsnames]{xcolor}

\usepackage{amsmath}
\usepackage{amssymb}
\usepackage{mathtools}
\usepackage{amsthm}
\allowdisplaybreaks[4]

\usepackage{pifont}
\usepackage{multirow}
\usepackage{colortbl}
\usepackage{threeparttable}
\usepackage{svg}
\usepackage{wrapfig}
\usepackage{graphicx}
\usepackage{subcaption}
\usepackage{cleveref} 

\usepackage{algorithm}
\usepackage{algorithmic}

\usepackage{bbding}

\usepackage{enumitem}

\newcommand\M{COSE}

\title{Compose Yourself: Average-Velocity Flow Matching for One-Step Speech Enhancement}

\name{Gang Yang$^{\dagger}$ \quad Yue Lei$^{\dagger}$ \quad Wenxin Tai$^{\dagger}$ \quad Jin Wu$^{\dagger}$ 
\quad Jia Chen$^{\dagger}$ \quad Ting Zhong$^{\dagger}$ \quad Fan Zhou$^{\dagger}$\sthanks{Corresponding author.}}
  
  \address{
  $^{\dagger}$ University of Electronic Science and Technology of China, Chengdu, Sichuan, China
  }
\begin{document}
\ninept
\maketitle

\begin{abstract}
% Diffusion and flow matching models have achieved excellent quality in speech enhancement, but modeling marginal vector fields often induces curved trajectories, causing severe discretization errors in large-step ODE solvers. In this study, we propose \M\, which learns a self-supervised average-velocity field through interval-splitting consistency, thereby avoiding costly Jacobian-vector products. To complement this unidirectional interval supervision, we further introduce a Closed-Loop Consistency (CLC) constraint, formulated as an algebraic calibration mechanism that enforces zero displacement over any closed time loop. CLC explicitly aligns forward and backward trajectories, providing a teacher-free signal that mitigates long-horizon error accumulation and improves the stability and convergence of learning. Finally, we propose a Self-Conditioned Adaptive Refinement (SCAR) sampling strategy, which is plug-and-play and can be seamlessly integrated with one-step generation methods. This approach leverages the outstanding one-step generation as a high-quality prior to guide subsequent sampling steps, thereby improving stability and consistency. Extensive experiments demonstrate \M's superior one-step performance and robust multi-step generation, outperforming state-of-the-art generative methods.
Diffusion and flow matching (FM) models have achieved remarkable progress in speech enhancement (SE), yet their dependence on multi-step generation is computationally expensive and vulnerable to discretization errors. Recent advances in one-step generative modeling, particularly MeanFlow, provide a promising alternative by reformulating dynamics through average velocity fields. In this work, we present \M, a one-step FM framework tailored for SE. To address the high training overhead of Jacobian-vector product (JVP) computations in MeanFlow, we introduce a velocity composition identity to compute average velocity efficiently, eliminating expensive computation while preserving theoretical consistency and achieving competitive enhancement quality. Extensive experiments on standard benchmarks show that \M~delivers up to 5x faster sampling and reduces training cost by 40\%, all without compromising speech quality. Code is available at \url{https://github.com/ICDM-UESTC/\M}.

\end{abstract}
\section{Introduction}
% 大纲换一下，介绍se，（分段） 然后转到生成式模型，diffusion与flow，然后说面临采样效率不高的挑战。（分段）
% mean-flow改善了效率，however（重要转折！）mean-flow需要复杂的微分操作，降低了可用性。-》因此，我们重新审视平均速度，基于积分中值定理拆解平均速度省去了jvp操作，更进一步的，我们利用XXX理论，对速度向量进行一致性约束，提升效率
% 之前的模型需要复杂的训练策略、多阶段、双模型、精细调度、预生成数据噪声对、提前停止等设计，否则会训练崩溃。我们提出一种简易的无需调度或warmup或jvp拟合训练目标的train from scratch一步流匹配模型用于语音增强，我们的方法类似于在训练期间执行自蒸馏，可以证明shortcut model是我们方法中时间步采样的一种特殊情况。
Speech enhancement (SE) aims to restore clean speech signals from recordings corrupted by noise, reverberation, and encoding artifacts~\cite{lim2005enhancement}. It benefits both human perception and downstream tasks such as automatic speech recognition (ASR)~\cite{li2021better}. Traditional methods based on statistical signal processing have been widely studied, but they often struggle to handle non-stationary or complex noise. The advent of deep learning has since driven significant progress by learning to directly map noisy speech to its clean counterpart~\cite{zheng2023sixty}. Recently, generative models~\cite{fu2021metricgan+,kang-etal-2025-llase, lu2022conditional} have gained prominence, excelling at capturing complex speech distributions and offering a powerful approach that preserves both speech quality and intelligibility.

Among generative models, diffusion and flow matching (FM) approaches have emerged as promising solutions for speech enhancement, demonstrating strong robustness to unseen noise~\cite{lu2022conditional} and are considered promising solutions for speech enhancement~\cite{yen2023cold,richter2023speech,guo2024variance,gonzalez2024investigating,tai2024dose}. Diffusion models use a forward stochastic differential equation (SDE) to transform clean speech, FM models directly learn a deterministic velocity field defined by an ordinary differential equation (ODE). Despite their impressive performance, these models still face challenges. Errors from large-step discretization degrade few-step performance, and both approaches remain computationally costly, often requiring five or more function evaluations (NFEs) during sampling~\cite{korostik2025modifying,lee2025flowse,li2024locally}.

Currently, a growing body of work has attempted to bridge this efficiency gap through
one-step generation techniques in FM~\cite{frans2025one,geng2025mean}. These methods avoid the error accumulation inherent in multi-step generation~\cite{yang2024consistency} and are considered promising for improving computational efficiency. Among these, the recently proposed MeanFlow framework~\cite{geng2025mean} is particularly notable. It provides an elegant mathematical formulation of generative dynamics by introducing the concept of average velocity over a time interval. This approach enables direct one-step generation via learned velocity fields, which holds great promise for multi-step generative speech enhancement by significantly improving the efficiency of generation.

% These approaches avoid the error accumulation inherent in multi-step generation~\cite{yang2024consistency} and are regarded as promising solutions for enhancing the computational efficiency of flow matching.

% 目前，有一些单步生成的方法出现，XXX实现了，XXX实现了XX，能够解决上述问题，显示了promising的前景。

% 受到这些优点的启发，我们提出了\M,引入meanflow到se的framework，为平均速度场与语音增强流匹配之间建立联系，实现了单步语音增强。此外，我们从ode的性质出发，通过向量合成的方法计算平均速度，从而避免jvp计算，进而降低训练开销。同时，我们还揭示了向量合成是一种拉直flow的等价方案，align with shortcut\consist flow。实验证明我们的方案实现了1 step生成，并取得了3.02的pesq，不需要昂贵的jvp操作。

In this work, we propose \M~  (\textit{\textbf{C}ompose vel\textbf{O}city in \textbf{S}peech  \textbf{E}nhancement}), a framework that integrates Meanflow~\cite{geng2025mean} into SE, enabling efficient, one-step generation. Further, by leveraging the properties of ODEs, we compute the average velocity via the velocity composition identity, which avoids \textit{Jacobian–vector product (JVP)} calculations and substantially reduces training overhead. Specifically, we incorporate Meanflow into SE by modeling the average velocity between two points on a curved trajectory, enabling the model to generate clean speech in one step. Building on this, we exploit the uniqueness property of ODEs to decompose displacements into compositions of two segment velocities, thereby eliminating the extra cost of direct \textit{JVP} computations. Moreover, we show that velocity composition yields an solution equivalent to other one-step generation methods, consistent with the self-consistency principle underlying the trajectory properties of flow matching.

Our main contributions are:
(1) We introduce \M, a novel framework that integrates Meanflow into speech enhancement to enable efficient one-step generation.
(2) We analyze that \textit{JVP} computation incurs significant overhead. Instead, we leveraged ODE properties to calculate average velocity, effectively avoiding \textit{JVP} and its associated computational cost.
(3) Our extensive experiments on standard benchmarks demonstrate that \M\ achieves at least 5x faster sampling and reduces GPU memory and training time by 40\% compared to MeanFlow, while maintaining equivalent performance.

% To address these issues, we propose \M, an Adaptive MeanFlow model for Speech Enhancement. Unlike prior methods that model instantaneous velocities, \M models the average velocity between two timesteps. This allows \M to capture sample evolution over long timesteps, producing satisfactory results even with large-step ODEs. Furthermore, we introduce a novel Self-Conditioned Adaptive Refinement strategy. This adaptive approach leverages a high-quality initial one-step pre-sampling estimate as a refined conditioning input for subsequent multi-step sampling, enabling \M to improve performance in multi-step scenarios. Extensive experiments demonstrate \M's state-of-the-art one-step generation performance, and the Self-Conditioned Enhancement strategy enables \M to maintain or even improve performance in multi-step sampling.

\section{Preliminary}
\subsection{Speech Enhancement}
Speech enhancement (SE) improves intelligibility and quality by suppressing noise while preserving natural speech characteristics. In the single-channel setting, the noisy signal can be expressed as:
\begin{equation}
\boldsymbol{y} = \boldsymbol{x} + \boldsymbol{n},
\label{eq:prelim}
\end{equation}
where $\boldsymbol{x}$ is the clean speech and $\boldsymbol{n}$ is additive noise. The objective is to estimate $\boldsymbol{x}$ from $\boldsymbol{y}$.

\begin{figure*}[t!] % Note the '*' here for a wide figure
    \centering
    \includegraphics[width=0.8\textwidth]{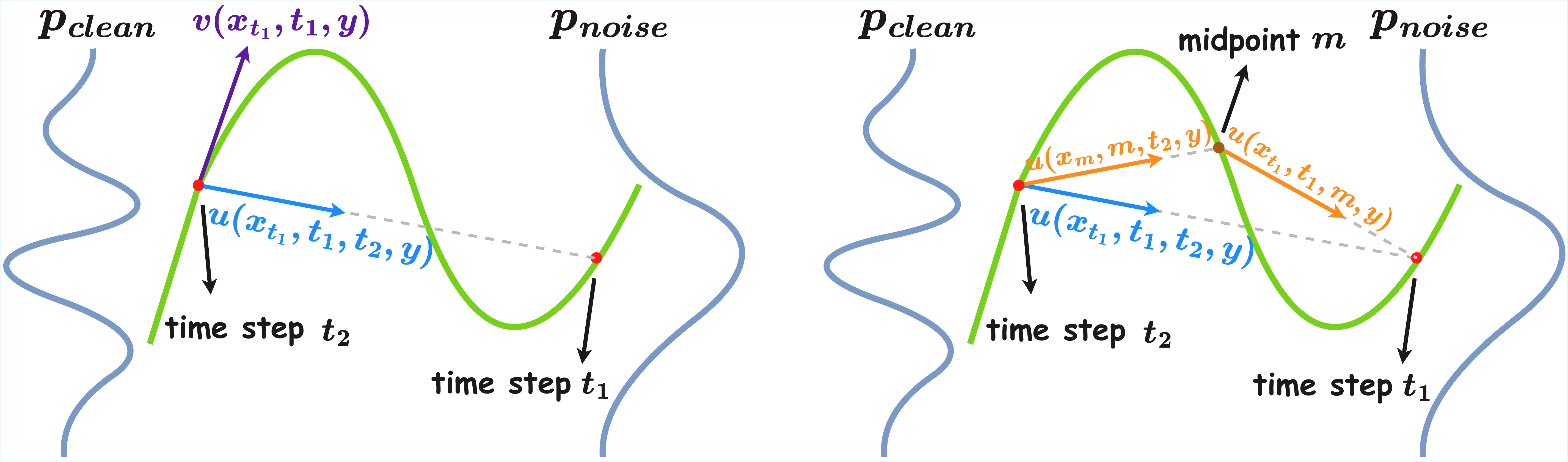} % Use a width relative to \textwidth to span both columns
    % \vspace{-0.2cm}
    \caption{Overview of velocity modeling. \textbf{Left}: Comparison between instantaneous and average velocity. The green line represents a generative trajectory between clean speech distribution and Gaussian noise distribution. \textbf{Right}: An example of velocity composition identity. For a midpoint $m$ between time steps $t_1$ and $t_2$, the two sub-average velocities can synthesize the average velocity of the large step.}
    \label{fig:workflow} % 给图添加一个标签，方便在文中引用
    % \vspace{-0.1cm}
\end{figure*}

\subsection{Flow Matching for Speech Enhancement}
Flow Matching (FM) models generative modeling as the learning of continuous-time velocity fields. It aims to learn a continuous-time trajectory $\boldsymbol{x}_t \in \mathbb{R}^d$, where $t \in [0, 1]$  from a simple distribution  to a complex clean speech distribution. A linear interpolation trajectory is commonly employed to construct intermediate states between a clean speech $\boldsymbol{x}_0$ and a sample from the prior $\boldsymbol{x}_1 \sim \mathcal{N}(0, \boldsymbol{\mathrm{I}})$~\cite{Liu2023flowstraight}:
\begin{equation}
\boldsymbol{x}_t = (1 - t)\boldsymbol{x}_0 + t\boldsymbol{x}_1, \quad \boldsymbol{x}_1 \sim \mathcal{N}(0, \boldsymbol{\mathrm{I}}),
\label{eq:linear_path}
\end{equation}
which defines a conditional distribution $p_t(\boldsymbol{x}_t \mid \boldsymbol{x}_0)$ from which intermediate states can be sampled explicitly.
The evolution of $\boldsymbol{x}_t$ is governed by a time-dependent instantaneous velocity field $v_t(\boldsymbol{x}_t)$, described by the following ordinary differential equation (ODE):
\begin{equation}
\frac{d\boldsymbol{x}_t}{dt} = v_t(\boldsymbol{x}_t), \quad t \in [0, 1].
\label{eq:flow_ode}
\end{equation}
Then a neural network $v_\theta(\boldsymbol{x}_t, t, \boldsymbol{y})$ is trained to approximate the velocity field with the Conditional Flow Matching (CFM) objective~\cite{lipman2022flow}:
\begin{equation}
\mathcal{L}_{\text{CFM}}(\theta) = \mathbb{E}_{t \sim \mathcal{U}(0, 1),\, \boldsymbol{x}_0,\, \boldsymbol{x}_1 \sim \mathcal{N}(0, I)} \left\| v_\theta(\boldsymbol{x}_t, t, \boldsymbol{y}) - (\boldsymbol{x}_1 - \boldsymbol{x}_0) \right\|^2,
\end{equation}
where $\boldsymbol{x}_t$ is sampled from~\cref{eq:linear_path} and $\boldsymbol{y}$ represents the noisy speech.

During inference, \cref{eq:flow_ode} is solved numerically using the Euler method with trained $v_\theta(\boldsymbol{x}_t, t)$ to generate clean speech:
\begin{equation}
\boldsymbol{x}_{t-\Delta t} = \boldsymbol{x}_t - \Delta t \, v_\theta(\boldsymbol{x}_t, t, \boldsymbol{y}), \quad \forall t \in [0, 1],
\label{eq:flow_inference}
\end{equation}
where $\Delta t = 1/N$, and $N$ represents the number of function evaluations (NFEs). By iteratively solving ~\cref{eq:flow_inference}, the speech can be refined from the noisy version to the clean version.

\section{Methodology}
In this section, we extend MeanFlow to speech enhancement and introduce a velocity composition identity to avoid computationally expensive operations. A overview of the velocity modeling approach is illustrated in~\cref{fig:workflow}.

\subsection{MeanFlow for Speech Enhancement} 
% Under the flow matching framework, the model learns the continuous-time instantaneous velocity field to achieve a generative mapping from the prior distribution to the target distribution.The core goal of flow matching is to learn a desired velocity field that is averaged over all potentially generated trajectories. Even under the optimal transport, this average velocity field induces a curved trajectory. When the input is pure noise at $t=1$, the velocity field learned by the model will tend to push all noise samples toward a point that represents the average of the dataset, which will cause one-step generation to fail completely.
Some recent work~\cite{geng2025mean,frans2025one} has observed that although the interpolated trajectories are designed to be straight during training, FM models can only predict the marginal velocity based on the current state during inference, which will naturally cause the trajectory to bend and deviate from the original straight line, leading one-step generation to fail completely. 

To address this challenge, MeanFlow builds an explicit model of the average velocity between two points on curved trajectory~\cite{geng2025mean}, as illustrated in~\cref{fig:workflow}~left. 
% In order to leverage this advantage for speech enhancement, we first define a simple straight-line flow path according to \cref{eq:liner_path} as:
% \begin{equation}
% \boldsymbol{x}_{t_1} = (1-t)\boldsymbol{x}_0 + t\boldsymbol{\epsilon},
% \label{eq:arb_flow_path}
% \end{equation}
% where $\boldsymbol{x}_0$ is a data sample and $\epsilon$ is a time-depend Gaussian noise sample.
% Our approach is based on the principles of optimal transport~\cite{Liu2023flowstraight}, which define a simple straight-line path between a data sample $\boldsymbol{x}_0$ and a Gaussion noise sample $\epsilon$:
% $$\boldsymbol{x}_{t_1} = (1-t)\boldsymbol{x}_0 + \epsilon_t.$$
Specifically, we first define the displacement $k(\boldsymbol{x}_{t_1}, t_1, t_2, \boldsymbol{y})$ between two time steps $t_1$ and $t_2$ as the integral of the instantaneous velocity field over that time interval as:
\begin{equation}
k(\boldsymbol{x}_{t_1}, t_1, t_2, \boldsymbol{y}) = \int_{t_2}^{t_1}v(\boldsymbol{x}_{\tau}, \tau, \boldsymbol{y})d\tau,
\label{eq:displacement}
\end{equation}
where $\boldsymbol{x}_{t_1}$ represents intermediate states at time step $t$, $\boldsymbol{y}$ represents noisy speech, and $v(\boldsymbol{x}_{\tau}, \tau, \boldsymbol{y})$ represents instantaneous velocity. The average velocity $u(\boldsymbol{x}_{t_1}, t_1, t_2, \boldsymbol{y})$ is then defined as this displacement divided by the time interval:
\begin{equation}
u(\boldsymbol{x}_{t_1}, t_1, t_2, \boldsymbol{y}) \triangleq \frac{k(\boldsymbol{x}_{t_1}, t_1, t_2, \boldsymbol{y})}{t_1-t_2} = \frac{1}{t_1-t_2}\int_{t_2}^{t_1}v(\boldsymbol{x}_{\tau}, \tau, \boldsymbol{y})d\tau.
\label{eq:mean_flow_def}
\end{equation}
By differentiating both sides of the equation with respect to $t_1$ simultaneously, this identity can be transformed into:
\begin{equation}
u(\boldsymbol{x}_{t_1}, t_1, t_2, \boldsymbol{y}) = v(\boldsymbol{x}_{t}, t_1, \boldsymbol{y}) - (t_1-t_2)\frac{d}{dt_1}u(\boldsymbol{x}_{t_1}, t_1, t_2, \boldsymbol{y}),
\label{eq:mean_flow_identity}
\end{equation}
where $\frac{d}{dt_1}u(\boldsymbol{x}_{t_1}, t_1, t_2, \boldsymbol{y})$ denotes the total derivative of the average velocity function $u$ with respect to time $t_1$.

According to~\cref{eq:mean_flow_identity}, we can train a neural network $u_\theta(\boldsymbol{x}_{t_1}, t_1, t_2, \boldsymbol{y})$ to directly model this average velocity field. The training objective encourages the network to satisfy the MeanFlow Identity:
\begin{equation}
\mathcal{L}(\theta)=\mathbb{E}_{\boldsymbol{x}_0, \boldsymbol{y}, t_1, t_2} \left\|u_{\theta}(\boldsymbol{x}_{t_1}, t_1, t_2, \boldsymbol{y}) - \text{sg}\left(u_{\text{tgt}}\right)\right\|_2^2,
\label{eq:mf_loss}
\end{equation}
where $u_{\text{tgt}} = v(\boldsymbol{x}_{t_1}, t_1,\boldsymbol{y}) - (t_1-t_2)\frac{d}{d{t_1}}u_{\theta}(\boldsymbol{x}_{t_1}, t_1, t_2, \boldsymbol{y})$. Following~\cite{geng2025mean}, we adopt the stop-gradient operation, denoted as $\text{sg}(\cdot)$, and replace $v(\boldsymbol{x}_{t_1}, t_1, \boldsymbol{y})$ with $\boldsymbol{x}_1 - \boldsymbol{x}_0$.  

In~\cref{eq:mf_loss}, $\frac{d}{dt_1}u_{\theta}(\boldsymbol{x}_{t_1}, t_1, t_2, \boldsymbol{y})$ is computed as the \textit{Jacobian–vector product (JVP)} between $[\partial_x u, \partial_{t_1} u, \partial_{t_2} u, \partial_{y} u]$ and the tangent vector $[v, 1, 0, 0]$, which can be implemented via PyTorch’s \textit{JVP} function. However, MeanFlow’s reliance on \textit{JVP} incurs significant overhead, limiting training efficiency.

% \subsection{Mean Flows with Interval-Splitting Consistency}
\subsection{Revisiting ODE Properties for Efficient Training}
\noindent\textbf{Computational Overhead of JVP.}
% We note that functions such as `torch.func.jvp` exhibit significant memory overhead. This is caused by a dual memory cost: its forward-mode automatic differentiation requires storing extra tangent values with each activation, and it simultaneously builds a backward graph for backpropagation. 
% % This overhead scales with data size. For the time-frequency domain model, under our experimental setup, it will result in an input size of $256*256$, thereby causing significant overhead.
The \textit{JVP} computes the directional derivative of a function along a given tangent vector. Modern automatic differentiation frameworks, such as PyTorch~\cite{paszke2017automatic}, use dual numbers to efficiency implement it as: 
\vspace{-1em}
\begin{align*}
u(\boldsymbol{x}_{t_1} +\boldsymbol{v} \delta,\, t_1+\delta,\, t_2,\, \boldsymbol{y}) 
= u(\boldsymbol{x}_{t_1}, t_1, t_2, \boldsymbol{y}) 
+ \left(\frac{\partial u}{\partial \boldsymbol{x}_{t_1}}\!\cdot\!\boldsymbol{v} + \frac{\partial u}{\partial t_1}\right) \!\delta,
\end{align*}
 where \( \delta \) is an infinitesimal for differentiation, and \( u(\boldsymbol{x}_{t_1}, t_1, t_2, \boldsymbol{y}) \) corresponds to the forward output, and the second term corresponds to the \textit{JVP} result. This approach requires only a single forward pass and theoretically introduces a slight computational cost. However, in practice, maintaining dual-numbered computations incurs memory allocations, additional arithmetic operations, and expanded computation graphs, leading to noticeable computational and memory costs~\cite{baydin2018automatic}. Furthermore, \textit{JVP} implementations differ across frameworks (e.g., PyTorch vs. JAX), adding development complexity and reducing portability. To address these limitations, we revisit the MeanFlow training method and propose a general solution that entirely circumvents \textit{JVP}, achieving efficient and memory-friendly training.

% We note that there exists efficient method for computing the JVP in modern libraries, which utilizes dual numbers~\cite{baydin2018automatic}. The core idea is to compute the \textit{JVP} in a single forward pass by replacing the function's input with a dual number, as~\cite{baydin2018automatic}'s (5).

% Specifically, for the function $u(\boldsymbol{x}, t_1, t_2, \boldsymbol{y})$ and the tangent vector $[\boldsymbol{v}, 1, 0, 0]$, dual numbers are applied to the input variables that have a non-zero component in the tangent vector. The dual part of the resulting output is the JVP, as follows::
% \begin{align*}
% u(\boldsymbol{x}+\boldsymbol{v}\epsilon, t_1+\epsilon, t_2, \boldsymbol{y})&=u(\boldsymbol{x}, t_1, t_2, \boldsymbol{y})+\left(\frac{\partial u}{\partial \boldsymbol{x}}\cdot\boldsymbol{v}+\frac{\partial u}{\partial t_1}\right)\epsilon
% \end{align*}

% Although this approach does not theoretically incur additional overhead, in practice it does, due to the large number of temporary variables created by modern implementations of automatic differentiation libraries for backward computation graphs. In addition, there is a clear mismatch between different library implementations, hence
% We revisit the MeanFlow training method to seek a general solution that can completely circumvent the JVP operation and achieve efficient training.

\noindent\textbf{Velocity Composition of ODE.} 
Inspired by previous work~\cite{frans2025one, yang2024consistency}, we derive a more general training framework for one-step generative models from the fundamental perspective of the uniqueness of ODE solutions. This framework allows us to bypass costly \textit{JVP} computations. Specifically, the ODE describes the deterministic evolution of a system. In line with previous studies~\cite{lipman2022flow,li2023self}, we assume that the instantaneous velocity field is locally Lipschitz continuous in the region traversed by the generated trajectories, denoted as $\mathcal{X}_t$:
\begin{equation}
\forall \boldsymbol{x}_1, \boldsymbol{x}_2 \in \mathcal{X}_t, \quad
\|v(\boldsymbol{x}_1, \tau, \boldsymbol{y}) - v(\boldsymbol{x}_2, \tau, \boldsymbol{y})\| \le L \|\boldsymbol{x}_1 - \boldsymbol{x}_2\|,
\end{equation}
where $\mathcal{X}_t$ represents the set of states visited by the ODE solution during generation.
% the ODE can then be rewritten as an integral equation whose solution can be obtained through Picard iteration~\cite{teschl2012ordinary}:
% \begin{equation}
%     \label{eq:picard_iteration}
%     \boldsymbol{x}_{\tau}^{(k+1)} = \boldsymbol{x}_{t_2} + \int_{r}^{\tau} v(\boldsymbol{x}_{\sigma}^{(k)}, \sigma, \boldsymbol{y}) d\sigma, \quad \tau \in [r, t].
% \end{equation}
% Under the Lipschitz condition,~\cref{eq:picard_iteration}
% % ~\cref{eq:picard_iteration} becomes a contraction mapping. According to the Banach fixed-point theorem~\cite{teschl2012ordinary},
% this mapping admits a unique and stable fixed point, thereby ensuring the uniqueness of the ODE solution. 
% Moreover, this result endows the ODE with the semigroup property: the evolution from $r$ to $t$ can be decomposed into a two-step process, first from $r$ to $s$ and then from $s$ to $t$, formalized as
Under this Lipschitz condition, the ODE satisfies the semigroup property~\cite{teschl2012ordinary}: its evolution from time $t_2$ to $t_1$ can be decomposed into a two-step process, first from $t_2$ to an intermediate time $m$, and then from $m$ to $t_1$. This can be formalized as:
\begin{align}
\Phi_{t_2 \to t_1} &= \Phi_{m \to t_1} \circ \Phi_{t_2 \to m}, \quad
s = t_2 + \alpha (t_1-t_2),
\end{align}
where $\Phi_{m \to t_1}$ is the evolution operator, namely $\Phi_{t_2 \to t_1}(\boldsymbol{x}_{t_2}) = \boldsymbol{x}_{t_1}$ and $\alpha \in [0,1]$, which maps the state at time $t_2$ to its corresponding state at time $t_1$ along the ODE trajectory. Consequently, the total displacement can be expressed as the sum of two sub-displacements:
\begin{equation}
    \label{eq:displacement_identity}
    k(\boldsymbol{x}_{t_1}, t_1, t_2, \boldsymbol{y}) = k(\boldsymbol{x}_{t_1}, t_1, m, \boldsymbol{y}) + k(\boldsymbol{x}_m, m, t_2, \boldsymbol{y}),
\end{equation}
According to the definition in~\cref{eq:mean_flow_def} and ~\cref{eq:displacement_identity}, we can further transform them into the following interpolation identity:
\begin{equation}
\label{eq:affine_interpolation}
\begin{split}
u(\boldsymbol{x}_{t_1}, t_1, t_2, \boldsymbol{y}) &= u(\boldsymbol{x}_m, m, t_2, \boldsymbol{y}) \\
\quad +& \alpha (u(\boldsymbol{x}_{t_1}, t_1, m, \boldsymbol{y}) - u(\boldsymbol{x}_m, m, t_2, \boldsymbol{y})),
\end{split}
\end{equation}
which can be interpreted as the velocity composition identity, combining two velocities into one, as shown in~\cref{fig:workflow} right. Here, the parameter $\alpha$ is randomly sampled from $[0,1]$, encouraging the model to learn consistency across different temporal granularities during training. Based on~\cref{eq:affine_interpolation}, we train the loss fuction of \M~without requiring expensive \textit{JVP} operations as:
\begin{equation}
\mathcal{L}_{\text{\M}}(\theta) = \mathbb{E}_{t,\boldsymbol{x}_t,\boldsymbol{y}} \left\| u_{\theta}(\boldsymbol{x}_{t_1}, t_1, t_2, \boldsymbol{y}) - \text{sg}\left(u_{\text{tgt}}\right) \right\|_2^2,
\end{equation}
where $\text{sg}(\cdot)$ denotes the stop-gradient operator following~\cite{geng2025mean} and $u_{\text{tgt}}$ defined as:
\begin{equation}
u_{\text{tgt}} = u_{\theta}(\boldsymbol{x}_m, m, t_2, \boldsymbol{y}) + \alpha(u_{\theta}(\boldsymbol{x}_{t_1}, t_1, m, \boldsymbol{y}) - u_{\theta}(\boldsymbol{x}_m, m, t_2, \boldsymbol{y})).
\nonumber
\end{equation}
The overall training procedure is summarized in ~\cref{alg:training}.

Beyond this, we establish a connection between velocity composition identity and self-consistency loss functions, ensuring the learning of a self-consistent and unidirectional average velocity field. Specifically, we divide the time interval \( [t_2, t_1] \) into two steps based on intermediate point \( m \) and let \( d_1 = t_1 - m \) and \( d_2 = m - t_2 \) represent the time intervals between \( t_1 \), \( m \), and \( t_2 \). Based on this, ~\cref{eq:affine_interpolation} can be transformed into:
\begin{equation}
\begin{split}
u_{\theta}(\boldsymbol{x}_{t_1},t_1,t_2,\boldsymbol{y}) &= \alpha \cdot u(\boldsymbol{x}_{t_1}, t_1, t_1 - d_1, \boldsymbol{y}) \\
\quad +& (1 - \alpha) \cdot u(\boldsymbol{x}_{t_1 - d_1}, t_1 - d_1 - d_2, \boldsymbol{y}), \\
% \text{where} \quad \boldsymbol{x}{t-d_1} &= \boldsymbol{x}_{t_1} - u_{\theta}(\boldsymbol{x}_{t_1}, t, d_1, \boldsymbol{y}) d.
\end{split}
\label{eq:affline_transformedtod}
\end{equation}
where $ \boldsymbol{x}_{t_1-d_1} = \boldsymbol{x}_{t_1} - u_{\theta}(\boldsymbol{x}_{t_1}, t_1, d_1, \boldsymbol{y}) $.
% We observe that when \( \alpha = \frac{1}{2} \), which results in \( d_1 = d_2 = d \),~\cref{eq:affline_transformedtod} can be specialized as:
% \begin{equation}
% \begin{split}
% u_{\theta}^{\text{S}}(\boldsymbol{x}_{t_1}, t_1, 2d, \boldsymbol{y}) &= u_{\theta}^{\text{S}}(\boldsymbol{x}_{t_1}, t_1, d, \boldsymbol{y}) / 2 \\
% &\quad + u_{\theta}^{\text{S}}(\boldsymbol{x}_{t_1-d}, t_1-d, d, \boldsymbol{y}) / 2.
% \end{split}
% \label{eq:shortcut}
% \end{equation}
% where $\boldsymbol{x}_{t_1-d}$ satisfied as
% $\boldsymbol{x}_{t_1-d} = \boldsymbol{x}_{t_1} - u_{\theta}^{\text{S}}(\boldsymbol{x}_{t_1}, t_1, d, \boldsymbol{y}) d.$
% We can observe that~\cref{eq:affline_transformedtod} aligns algebraically with the self-consistency identity in \cite{frans2025one}'s (4). According to \cite{yang2024consistency}, this identity enables propagating generation capabilities across step scales (multi→few→one) while supporting joint training of the unified objective via a single model in an end-to-end run. Unlike counterparts differing only in conditional time parameterization \cite{geng2025mean}, \M\ offers greater flexibility as a universal framework accommodating arbitrary time intervals and evolution patterns.
We observe that~\cref{eq:affline_transformedtod} is algebraically equivalent to the self-consistency identity in ~\cite[eq.~(4)]{frans2025one}, which enables propagating generation across step scales (multi→few→one) and joint end-to-end training~\cite{yang2024consistency}. \M\ builds on this principle to provide a flexible and general framework that accommodates arbitrary time intervals and diverse evolution patterns.

\begin{algorithm}[t!]
\caption{\M\ Training Algorithm}
\label{alg:training}
\begin{algorithmic}[1]
\REQUIRE Neural network $u_\theta$, a batch of clean data $x$ and noisy data $\boldsymbol{y}$, optimizer.
\STATE Sample time points $t_2, t_1$ such that $0 \le t_2 \le t_1 \le 1$.
\STATE Sample $\alpha \sim \mathcal{U}(0, 1)$, set $m = t_1 + \alpha(t_2 - t_1)$.
\STATE Sample prior $\epsilon \sim \mathcal{N}(0, I)$.
\STATE Construct flow path point at time $t$: $\boldsymbol{x}_{t_1} = (1-t_1)x + t_1\epsilon$.
\STATE $u_2 = u_\theta(\boldsymbol{x}_{t_1}, t_1, m, \boldsymbol{y})$, $\boldsymbol{x}_m = \boldsymbol{x}_{t_1} - (t_1-m)u_2$.
\STATE $u_1 = u_\theta(\boldsymbol{x}_m, m, t_2, \boldsymbol{y})$, $\boldsymbol{x}_{t_2} = \boldsymbol{x}_m - (m-t_2)u_1$.
\STATE $u_{t_1t_2} = u_\theta(\boldsymbol{x}_{t_1}, t_1, t_2, \boldsymbol{y})$, $u_{\text{tgt}} = u_1 + \alpha(u_2 - u_1)$
\STATE Update $\theta$ via $\nabla_\theta \mathcal{L}_{\text{\M}}, \mathcal{L}_{\text{\M}} = \| u_{t_1t_2} - \text{sg}(u_{\text{tgt}}) \|_2^2$

\end{algorithmic}
\end{algorithm}

\section{Experiments}

\begin{table*}[ht]
\centering
\vspace{-0.2cm}
\small % 或者 \footnotesize，根据需要调整字体大小
\caption{Speech enhancement results on VoiceBank-DEMAND and ChiME-4 datasets. All experimental results are presented as mean $\pm$ standard deviation. The best results are highlighted in bold.}
% \vspace{-0.2cm}
\label{tab:performance_comparison}
\resizebox{\linewidth}{!}{
% \begin{tabular}{p{1.25cm} c | p{1cm} p{1cm} p{1.15cm} p{1.15cm} p{1.15cm} | p{1cm} p{1cm} p{1.15cm} p{1.15cm} p{1.15cm}} % 将 'l' 改为 'p{2.5cm}'，并移除了 @{\extracolsep{\fill}}
\begin{tabular}{p{1.25cm} c | c c c c c | c c c c c} % 将 p{...} 改为 l 和 c，以实现居中
\toprule
\midrule
\multirow{2}{*}{Method} & \multirow{2}{*}{NFE} & \multicolumn{5}{c}{VoiceBank-
DEMAND} & \multicolumn{5}{c}{CHiME-4} \\
\cmidrule(lr){3-7} \cmidrule(lr){8-12}
& & PESQ & ESTOI & SI-SDR & SI-SIR & SI-SAR & PESQ & ESTOI & SI-SDR & SI-SIR & SI-SAR \\
\midrule
Mixture & - & 1.97$^{\pm0.75}$ & 0.79$^{\pm0.15}$ & 8.4$^{\pm5.6}$ & 8.5$^{\pm5.6}$ & 47.5$^{\pm10.4}$ & 1.27$^{\pm0.16}$ & 0.68$^{\pm0.08}$ & 7.5$^{\pm2.1}$ & 7.5$^{\pm2.1}$ & 46.7$^{\pm10.3}$ \\
% CDiffuSE的chime4
% CDiffuSE & 200 & 2.52$^{\pm0.58}$ & 0.79$^{\pm0.10}$ & 12.4$^{\pm2.8}$ & 19.8$^{\pm6.0}$ & 13.8$^{\pm1.8}$ & 2.52$^{\pm0.58}$ & 0.79$^{\pm0.10}$ & 12.4$^{\pm2.8}$ & 19.8$^{\pm6.0}$ & 13.8$^{\pm1.8}$ \\
NCSN++~\cite{richter2023speech} & 1 & 2.87$^{\pm0.74}$ & \textbf{0.87$^{\pm0.10}$} & 19.1$^{\pm3.5}$ & 31.5$^{\pm7.2}$ & 20.0$^{\pm3.5}$ & 1.27$^{\pm0.16}$ & 0.68$^{\pm0.08}$ & 7.4$^{\pm2.1}$ & 7.4$^{\pm2.1}$ & \textbf{42.4$^{\pm5.4}$} \\
SGMSE+~\cite{richter2023speech} & 15 & 2.80$^{\pm0.58}$ & 0.86$^{\pm0.10}$ & 17.2$^{\pm3.6}$ & 26.9$^{\pm5.2}$ & 17.9$^{\pm3.5}$ & \textbf{1.82$^{\pm0.29}$} & 0.82$^{\pm0.07}$ & 13.5$^{\pm2.3}$ & 25.2$^{\pm3.4}$ & 13.9$^{\pm2.2}$ \\
SGMSE+~\cite{richter2023speech} & 5& 1.15$^{\pm0.08}$ & 0.63$^{\pm0.10}$ & 7.9$^{\pm1.8}$ & 18.6$^{\pm4.1}$ & 8.4$^{\pm1.6}$ & 1.09$^{\pm0.04}$& 0.58$^{\pm0.09}$ & 5.3$^{\pm1.6}$ & 15.2$^{\pm2.1}$& 5.8$^{\pm1.6}$\\
SGMSE+~\cite{richter2023speech} & 1 & 1.06$^{\pm0.10}$ & 0.02$^{\pm0.02}$ & $-$25.3$^{\pm1.7}$ & 23.3$^{\pm9.9}$ & $-$25.3$^{\pm1.7}$ & 1.08$^{\pm0.14}$ & 0.02$^{\pm0.02}$ & $-$26.6$^{\pm1.2}$ & 22.8$^{\pm8.9}$ & $-$26.6$^{\pm1.2}$ \\
StoRM~\cite{lemercier2023storm} & 15& 2.77$^{\pm0.57}$ & \textbf{0.87}$^{\pm0.10}$ & 18.5$^{\pm3.3}$ & 30.9$^{\pm6.7}$ & 19.1$^{\pm3.3}$ & \textbf{1.82$^{\pm0.30}$} & \textbf{0.84$^{\pm0.06}$} & $14.5^{\pm2.2}$ & $26.0^{\pm3.8}$ & \textbf{$14.8^{\pm2.2}$}\\
StoRM~\cite{lemercier2023storm} & 5& 1.25$^{\pm0.09}$ & 0.70$^{\pm0.10}$ & 11.5$^{\pm1.4}$ & 29.8$^{\pm6.1}$ & 11.6$^{\pm1.4}$ & $1.19^{\pm0.08}$ & \textbf{$0.72^{\pm0.08}$} & $9.9^{\pm1.6}$ & $24.5^{\pm3.3}$ & \textbf{$10.0^{\pm1.6}$}\\
StoRM~\cite{lemercier2023storm} & 1& 1.04$^{\pm0.02}$ & 0.10$^{\pm0.03}$ & $-$16.9$^{\pm1.5}$ & 22.2$^{\pm7.0}$ & $-$16.9$^{\pm1.5}$ & $1.05^{\pm0.12}$ & $0.10^{\pm0.03}$ & $-$17.7$^{\pm1.1}$ & 18.3$^{\pm3.4}$ & $-$17.7$^{\pm1.1}$ \\
VPIDM~\cite{guo2024variance} & 15 & 2.92$^{\pm0.61}$ & \textbf{0.87$^{\pm0.10}$} & 18.8$^{\pm2.5}$ & 28.4$^{\pm5.3}$ & 19.4$^{\pm1.9}$& $1.67^{\pm0.20}$ & 0.80$^{\pm0.07}$ & 14.2$^{\pm1.9}$ & $23.2^{\pm2.9}$ & 13.5$^{\pm2.9}$\\
VPIDM~\cite{guo2024variance} & 5 & 1.85$^{\pm0.12}$ & 0.83$^{\pm0.10}$ & 14.6$^{\pm2.1}$ & 17.7$^{\pm3.9}$ & 14.9$^{\pm1.6}$& $1.03^{\pm0.01}$ & 0.33$^{\pm0.07}$ & $-$6.7$^{\pm1.5}$ & $12.2^{\pm1.8}$ &$-$6.6$^{\pm1.5}$\\
VPIDM~\cite{guo2024variance} & 1 & 1.04$^{\pm0.01}$ & 0.29$^{\pm0.06}$ & $-$7.2$^{\pm1.5}$ & 14.3$^{\pm3.7}$ & $-$7.2$^{\pm1.5}$ & $1.04^{\pm0.01}$ & 0.29$^{\pm0.06}$ & $-$7.2$^{\pm1.5}$ & $14.3^{\pm3.7}$ &$-$7.2$^{\pm1.5}$\\
\midrule
FlowSE~\cite{lipman2022flow} & 5 & 2.96$^{\pm0.73}$& \textbf{0.87}$^{\pm0.10}$ & 18.8$^{\pm3.5}$ & \textbf{31.7$^{\pm7.0}$}& 19.8$^{\pm3.5}$& 1.73$^{\pm0.31}$& \textbf{0.84$^{\pm0.06}$}& 14.2$^{\pm2.3}$& 24.3$^{\pm3.5}$& 14.7$^{\pm2.3}$\\
FlowSE~\cite{lipman2022flow} & 1 & 1.37$^{\pm0.15}$& 0.74$^{\pm0.10}$ & 12.1$^{\pm2.0}$ & 22.2$^{\pm4.6}$& 12.7$^{\pm1.7}$& 1.06$^{\pm0.03}$& 0.53$^{\pm0.09}$ & 2.0$^{\pm1.5}$ & 16.1$^{\pm2.3}$& 2.1$^{\pm1.5}$\\
LARF~\cite{li2024locally} & 5 & 2.98$^{\pm0.75}$ & \textbf{0.87$^{\pm0.10}$} & 18.8$^{\pm3.4}$ & 26.6$^{\pm7.4}$ & 20.2$^{\pm3.5}$ & 1.67$^{\pm0.27}$ & \textbf{0.84$^{\pm0.07}$} & \textbf{14.8$^{\pm2.4}$} & 25.7$^{\pm3.8}$ & 15.2$^{\pm2.3}$\\
LARF~\cite{li2024locally} & 1 & 2.97$^{\pm0.70}$ & \textbf{0.87$^{\pm0.10}$} & 19.2$^{\pm3.7}$ & 26.4$^{\pm5.6}$ & \textbf{20.7$^{\pm3.7}$} & 1.66$^{\pm0.28}$ & 0.83$^{\pm0.07}$ & 14.5$^{\pm2.3}$ & 25.6$^{\pm3.8}$ & 15.5$^{\pm2.3}$\\
\midrule\midrule
\M & 1 & \textbf{3.02$^{\pm0.74}$}& \textbf{0.87$^{\pm0.10}$} & \textbf{19.3$^{\pm3.4}$} & \textbf{31.7$^{\pm6.2}$} & 19.8$^{\pm3.5}$ & 1.76$^{\pm0.29}$ & \textbf{0.84$^{\pm0.07}$} & 14.3$^{\pm2.5}$ & \textbf{26.1$^{\pm3.7}$} & 14.6$^{\pm2.4}$\\ % 占位符
\midrule
\bottomrule
\end{tabular}
}
% \vspace{-0.3cm}
\end{table*}

\begin{table}[t]
\centering
\caption{Performance and training overhead of one-step generation.}
\label{tab:ours vs meanflow}
\resizebox{\linewidth}{!}{% % \linewidth 表示当前行的宽度，! 保持纵横比
\begin{tabular}{ccccccc} % 调整了每列的宽度
\toprule
Method  & PESQ & ESTOI & SI-SDR & Time (ms/step) & Memory \\
\midrule
NCSN++ &2.87$^{\pm0.74}$ & 0.87$^{\pm0.10}$ & 19.1$^{\pm3.5}$ & 70 & 3733\\
FlowSE & 1.25$^{\pm0.11}$& 0.71$^{\pm0.10}$ & 10.8$^{\pm2.0}$ & 70 & 3733\\
MeanFlow & 3.00$^{\pm0.73}$& 0.87$^{\pm0.10}$& 19.1$^{\pm3.4}$& 195 & 14005\\
\M\  & 3.02$^{\pm0.74}$& 0.87$^{\pm0.10}$ & 19.3$^{\pm3.4}$ & 112 & 8441 \\
\bottomrule
\end{tabular}%
} % resizebox 命令的结束
% \vspace{-0.5cm}
\end{table}

% \begin{table}
% \caption{Performance comparison of MeanFlow and \M\ for one-step generation on VBD dataset.}
% \label{tab:ours vs meanflow}
% \renewcommand{\arraystretch}{1.2} % 增加行高
% \centering % 居中表格
% \begin{tabular}{p{2cm} p{2cm} p{2cm}} % 手动调整列宽以适应页面
% \toprule
% Metric & MeanFlow & \M \\
% \midrule
% NFE & 1 & 1 \\
% PESQ & 3.00$^{\pm0.73}$ & 3.02$^{\pm0.74}$ \\
% ESTOI & 0.87$^{\pm0.10}$ & \textbf{0.87$^{\pm0.10}$} \\
% SI-SDR & 19.1$^{\pm3.4}$ & 19.3$^{\pm3.4}$ \\
% ms/step & 195 & 112 \\
% Memory & 14005 & 8441 \\
% \bottomrule
% \end{tabular}
% \end{table}

\subsection{Experimental Settings}
\noindent\textbf{Datasets.} Our method is evaluated on the VoiceBank-DEMAND (VBD)~\cite{veaux2013voice} and CHiME-4~\cite{vincent2017analysis} datasets.   The VBD training set contains recordings from 28 speakers (14 female, 14 male) with 8 noise types. For testing, we use the VBD test set with two unseen speakers and two new noise types. For a robust cross-dataset evaluation, we follow \cite{lu2022conditional, tai2024dose} and use the fifth-microphone recordings from CHiME-4, which feature real-world noise from four environments: street, walking area, cafeteria, and bus.

\noindent\textbf{Implementation Details.}
 Following previous work~\cite{geng2025mean}, we extend the NCSN++ architecture with a Fourier-embedded channel and concatenated $(t_1, t_2)$ steps to model average velocity, with negligible parameter overhead. We also adopt the same training setup: sampling timestep from a $\mathrm{lognorm}(-0.4, 1.0)$, setting $t_2=t_1$ with 50\% probability, and using the adaptive L2 loss $\mathcal{L}=w|\Delta|_2^{2\gamma}$, where $w=1/(|\Delta|_2^2+c)^p$ and $c=10^{-3}$.

% Through extensive experimentation, we determined the $p$ value to be $0.5$, which makes the loss function numerically similar to the Pseudo-Huber loss. The final adaptively weighted loss expression is $\mathcal{L} = \text{sg}(w) \cdot \|\Delta\|_2^2$.

\noindent\textbf{Metrics.}
We quantitatively evaluate performance using several standard metrics: Perceptual Evaluation of Speech Quality (PESQ)~\cite{Rix2001Perceptual} for perceived quality, Extended Short-Time Objective Intelligibility (ESTOI)~\cite{jensen2016algorithm} for intelligibility, and the scale-invariant measures SI-SDR, SI-SIR, and SI-SAR~\cite{le2019sdr}.

\noindent\textbf{Configurations.}
We follow~\cite{richter2023speech} in using the same backbone and audio preprocessing, applying a 512-sample STFT with a 128-sample hop to obtain 256 frequency bins. Models are trained for 200 epochs on a single NVIDIA 4090 GPU with a learning rate of $5\times10^{-5}$.

\noindent\textbf{Baselines.}
% 这里还不够，比较的baseline太少了
We compare our proposed methodology with three baseline models:  
% CDiffuSE~\cite{lu2022conditional}, 
SGMSE+~\cite{richter2023speech}, StoRM~\cite{lemercier2023storm}, VPIDM~\cite{guo2024variance}, FlowSE~\cite{lee2025flowse,wang2025flowse}, LARF~\cite{li2024locally}. For FlowSE, we adopt the method from two concurrent conference works~\cite{lee2025flowse,wang2025flowse} and re-implement it following the diffusion schedule and architecture in~\cite{lipman2022flow,lee2025flowse}, as we did not reproduce the reported results in our setting using the available implementation. Following prior work~\cite{guo2024variance}, we also incorporate the NCSN++~\cite{richter2023speech} architecture as a discriminative model in our comparisons.
% Note that there are various other diffusion-based SE methods, such as DR-DiffuSE~\cite{tai2023revisiting}, SRTNet~\cite{qiu2023srtnet}, and UNIVERSE++~\cite{scheibler2024universal}. These methods use similar strategies to the chosen baselines but improve consistency through complicated architecture designs. Therefore, we believe the selected baselines are sufficient for evaluation.

\subsection{Experimental Results}
\noindent\textbf{Overall Performance.}
\Cref{tab:performance_comparison} summarizes \M's speech enhancement results on VBD and CHiME-4 datasets, compared with a range of diffusion-based and flow-matching baselines. In terms of overall performance, \M\ achieves superior one-step generation: on VBD, it outperforms diffusion models such as SGMSE+~\cite{richter2023speech}, StoRM~\cite{lemercier2023storm}, and VPIDM~\cite{guo2024variance} even when with 15 steps, and it also surpasses advanced flow-matching methods such as FlowSE~\cite{lee2025flowse,wang2025flowse,lipman2022flow} and LARF~\cite{li2024locally}. On CHiME-4, \M\ remains competitive, confirming its robustness under diverse acoustic conditions.

\Cref{tab:performance_comparison} also provides detailed results across different sampling steps for diffusion-based and FM baselines. It reveals that diffusion-based baselines maintain acceptable quality at 15 steps but degrade rapidly as the step count decreases, with one-step PESQ collapsing to around 1.0. Similarly, FlowSE~\cite{lipman2022flow}, which models instantaneous velocities, fails entirely in one-step generation because the large-step discretization leads to severe trajectory deviation. In contrast, \M\ consistently produces high-quality results in just one step and surpasses LARF~\cite{li2024locally}. These embody the ability of average velocity modeling to reduce cumulative errors and generate in one step.

\noindent\textbf{Ablation Study.} \Cref{tab:ours vs meanflow} summarizes the one-step performance and training overhead of different methods.The result shows that although MeanFlow is effective, it incurs substantial computational overhead: each training step takes 195ms and consumes 14,005MB of GPU memory, approximately 2.75× higher than that of NCSN++ and FlowSE, significantly limiting its practicality. In contrast, \M\ adopts the velocity composition identity in place of the MeanFlow identity, effectively avoiding the costly \textit{JVP} operations. This leads to a notable reduction in training overhead, with training time reduced by approximately 43\% and GPU memory usage by about 40\%, while further improving performance. Although \M\ has less training overhead than MeanFlow, its cost remains higher than NCSN++ due to two additional forward per training step.

\begin{figure}[!t]  
\begin{center}
    \includegraphics[width=\columnwidth]{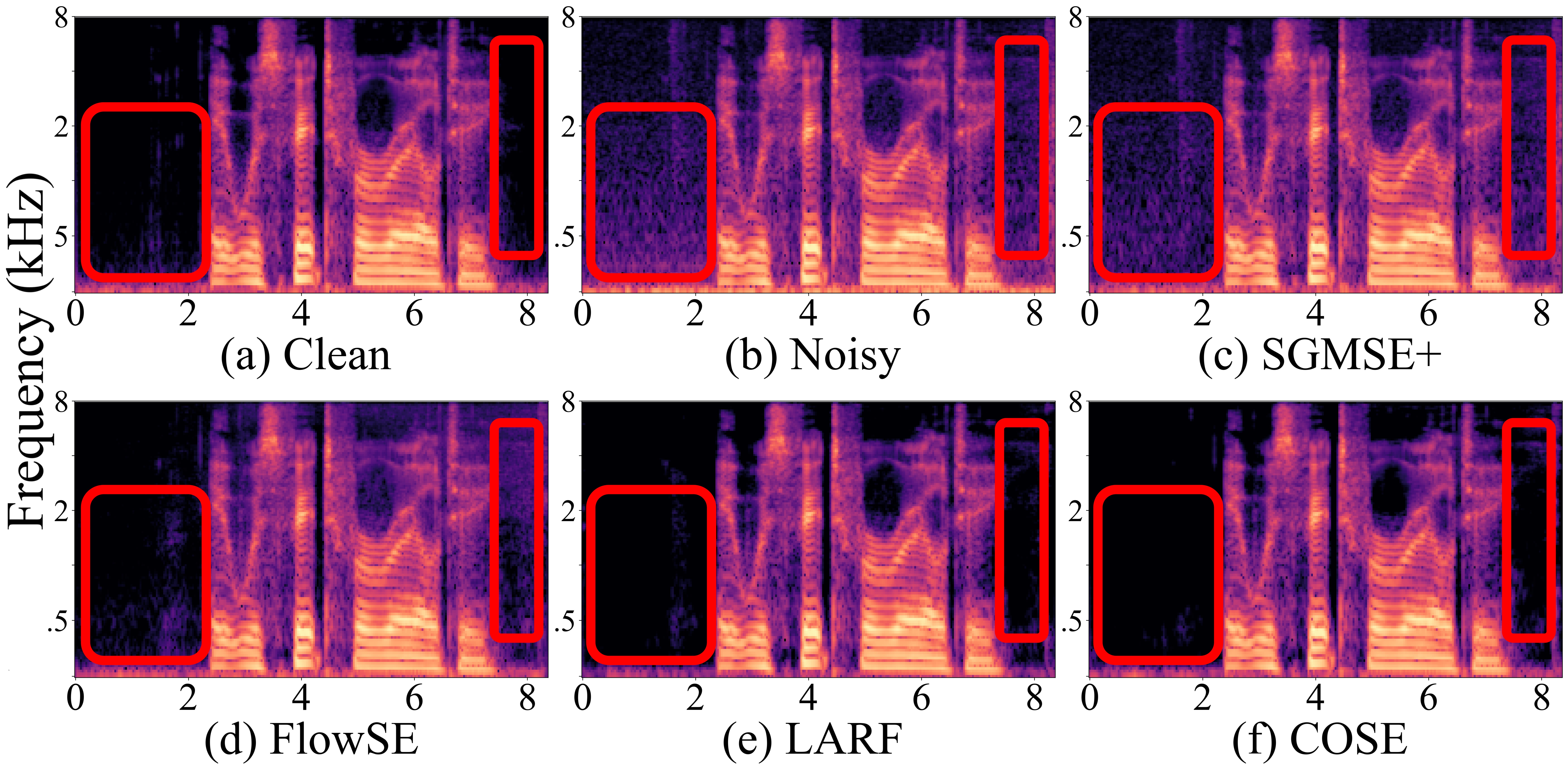}
    \vspace{-0.6cm}
    \caption{Visualization of magnitude spectrum.}
    \vspace{-0.9cm}
    \label{fig:case_study}
\end{center}
\end{figure}

\noindent\textbf{Case Study.} \Cref{fig:case_study} shows a visual case, where red rectangles demonstrate that \M\ outperforms other baseline models in noise removal while preserving speech details. This advantage originates from \M’s one-step generation capability, which avoids cumulative errors and improves generation quality.

\section{Conclusion}
In this work, we proposed \M, an one-step flow matching framework that aims to guide speech enhancement process with average velocity. Our method circumvents costly \textit{JVP} computations by leveraging a velocity composition identity. This approach significantly reduces training overhead and provides a solution equivalent to other one-step generation methods. Experiments on standard benchmarks demonstrate that \M\ achieves competitive enhancement quality with significantly improved efficiency, showing the promise of one-step flow matching for practical SE applications. We hope this work encourages further development of one-step generation in SE. In future work, we plan to explore the generalization of our approach across a broader range of model architectures and datasets. 

\clearpage
\ninept

\bibliographystyle{IEEEbib}
\bibliography{refs}

\end{document}